\documentclass[prb,twocolumn,superscriptaddress]{revtex4-1}
\usepackage[T1]{fontenc}
\usepackage[utf8]{inputenc}
\usepackage{graphicx}
\usepackage{amsmath}
\usepackage{color}
\usepackage{ulem}
\usepackage{comment}
\usepackage{hyperref}
\usepackage{caption}
\usepackage{subcaption}
\usepackage[left=1.5cm,right=1.5cm,top=2cm,bottom=2cm]{geometry}
\usepackage{epstopdf}
\begin{document}

\title{Coupled Cluster Method with Single and Double Excitations Tailored \\ by Matrix Product State Wave Functions}

\author{Libor Veis}
\email{libor.veis@jh-inst.cas.cz}
\affiliation{J. Heyrovsk\'{y} Institute of Physical Chemistry, Academy of Sciences of the Czech \mbox{Republic, v.v.i.}, Dolej\v{s}kova 3, 18223 Prague 8, Czech Republic}

\author{Andrej Antal\'{i}k}
\affiliation{J. Heyrovsk\'{y} Institute of Physical Chemistry, Academy of Sciences of the Czech \mbox{Republic, v.v.i.}, Dolej\v{s}kova 3, 18223 Prague 8, Czech Republic}

\author{Ji\v{r}\'{i} Brabec}
\affiliation{J. Heyrovsk\'{y} Institute of Physical Chemistry, Academy of Sciences of the Czech \mbox{Republic, v.v.i.}, Dolej\v{s}kova 3, 18223 Prague 8, Czech Republic}

\author{Frank Neese}
\affiliation{Max Planck Institut f\"{u}r Chemische Energiekonversion, Stiftstr. 34-36, D-45470 M\"{u}lheim an der Ruhr, Germany}

\author{\"Ors Legeza}
\email{legeza.ors@wigner.mta.hu}
\affiliation{Strongly Correlated Systems ``Lend\"{u}let'' Research group, Wigner Research Centre for Physics, H-1525, Budapest, Hungary}

\author{Ji\v{r}\'{i} Pittner}
\email{jiri.pittner@jh-inst.cas.cz}
\affiliation{J. Heyrovsk\'{y} Institute of Physical Chemistry, Academy of Sciences of the Czech \mbox{Republic, v.v.i.}, Dolej\v{s}kova 3, 18223 Prague 8, Czech Republic}

\date{\today}

\begin{abstract}
In the last decade, the quantum chemical version of the density matrix renormalization group (DMRG) method has established itself as the method of choice for calculations of strongly correlated molecular systems. Despite its favourable scaling, it is in practice not suitable for computations of dynamic correlation.
We present a novel method for accurate ``post-DMRG'' treatment of dynamic correlation based on the tailored coupled cluster (CC) theory in which the DMRG method is responsible for the proper description of non-dynamic correlation, whereas dynamic
 correlation is incorporated through the framework of the CC theory.
We illustrate the potential of this method on prominent multireference systems, in particular N$_2$, Cr$_2$ molecules and also oxo-Mn(Salen) for which we have performed the first ``post-DMRG'' computations in order to shed light on the energy ordering of the lowest spin states.
\end{abstract}


\maketitle

The coupled cluster (CC) approach, introduced to quantum chemistry (QC) by \v{C}\'{\i}\v{z}ek \cite{cizek-original}, is one of the most accurate \textit{ab initio} methods
for the treatment of dynamic electron correlation. The advantages of this scheme include a compact description of the wave function, size-extensivity, invariance to orbital rotations together with a systematic hierarchy of approximations converging towards the full configuration interaction (FCI) limit \cite{gauss-encyclop}. 
Despite the great success of QC and in particular the CC methodology \footnote{
The CCSD(T) method \cite{ccsdtparent1}, which includes connected single, double, and perturbative triple excitations, is often referred to  as gold standard of quantum chemistry.
} in standard (single-reference) cases, 
the situation is dramatically different for strongly correlated (multireference) systems \footnote{Many important compounds belong to this class, e.g. reaction intermediates or transition metal complexes.},
where the usual single-reference approaches become inaccurate or even completely break down.
One category of methods designed for the treatment of such systems are multireference coupled cluster (MRCC) approaches,
which generalize the CC exponential parameterization of the wave function \cite{bartlett-musial2007,ourspringer,bartlett_mrcc_2012}.
Out of many formulations of MRCC theories, the class of methods relevant to this work are externally corrected CC,
which extract information about the most important higher excitations or active space single and double excitations from an ``external'' calculation performed by a different method like complete active space
self-consistent field (CASSCF) or multireference configuration interaction (MRCI) \cite{paldus-externalcorr,paldus-externalcorr2,paldus-externalcorr-new1,paldus-externalcorr-new2,planelles1994a,tobola1996,peris1997,kinoshita_2005,cyclobut-tailored-2011,melnichuk-2012,melnichuk-2014,tcc-appl}.
In this letter, we present a further development in this field concerning the tailored CC (TCC) method, where
the information for external correction is obtained from a density matrix renormalization group (DMRG) calculation.

DMRG is a very powerful approach suitable for treatment of strongly correlated systems
originally 
developed 
in solid state physics \cite{White-1992a,White-1992b,White-1993}.
The success of DMRG in this field motivated its application to QC problems \cite{white_1999,Chan-2002a,legeza_2003a,Legeza-2008,marti_2010,chan_review,wouters_review,legeza_review,yanai_review}
where it has proven the potential to 
outperform traditional QC methods for systems which require very large 
active 
spaces, like 
molecules
containing
several transition metal atoms \cite{Kurashige-2013,sharma_2014b}.
Despite the favourable scaling of the DMRG method, it is computationally prohibitive to treat the dynamic correlation by including all virtual orbitals into the active space. Since the dynamic correlation has in general a very significant chemical impact, developement of ''post-DMRG'' methods, which aim to describe this effect, is of high importance.
During the past few years, several such methods have been developed, 
for example 
DMRG-CASPT2 \cite{kurashige_2011}, DMRG-icMRCI \cite{Saitow-2013}, Canonical Transformation (CT) \cite{neuscamman_2010_irpc}, or the matrix product state (MPS)-based formulation of a multireference perturbation theory \cite{Sharma-2014b}.

The general TCC wave function employs the following split-amplitude ansatz \cite{kinoshita_2005}

\begin{equation}
  \label{eq:TCC}
  | \Psi_\text{TCC} \rangle = e^{T} | \Phi_\text{0} \rangle =  e^{T_\text{ext}+T_\text{CAS}} | \Phi_\text{0} \rangle =  e^{T_\text{ext}} e^{T_\text{CAS}} | \Phi_\text{0} \rangle,
\end{equation}

\noindent
where $T_\text{CAS}$ represents the amplitudes obtained from the CI coefficients of the 
pre-computed complete active space configuration interaction (CASCI)
wave function and $T_\text{ext}$ is the rest of the cluster operator. 
Since $| \Phi_\text{0} \rangle$ is a single-determinant reference wave function, $T_\text{ext}$ and $T_\text{CAS}$ mutually commute, which keeps the method very simple.
At the level of truncation to single and double excitations (TCCSD), the wave function reads
\begin{equation}
  \label{eq:TCCSD}
  | \Psi_\text{TCCSD} \rangle = e^{\big(T^{(1)}_\text{ext} + T^{(2)}_\text{ext}\big)} e^{\big(T^{(1)}_\text{CAS} + T^{(2)}_\text{CAS}\big)} | \Phi_\text{0} \rangle,
\end{equation}

\noindent
where the superscript denotes the excitation rank of a cluster operator. $T^{(1)}_\text{CAS}$ and $T^{(2)}_\text{CAS}$ are calculated from the CASCI expansion coefficients according to the well-know relationship between the CC and CI expansions

\begin{subequations}
\begin{eqnarray}
	\label{eq:ci2cc}
	T^{(1)}_{\text{CAS}} & = & C^{(1)}, \\  
	T^{(2)}_{\text{CAS}} & = & C^{(2)} - \frac{1}{2}[C^{(1)}]^2.
\end{eqnarray}
\end{subequations}

\noindent
$T^{(1)}_{\text{CAS}}$ and $T^{(2)}_{\text{CAS}}$ are expected to properly describe the non-dynamic correlation\footnote{Notice also that due to the two-body nature of the electronic Hamiltonian, TCCSD energy within the active space will reproduce the CASCI energy.} and are kept constant during the CC procedure. 
They thus ``tailor'' the external amplitudes corresponding to the $T^{(1)}_{\text{ext}}$ and $T^{(2)}_{\text{ext}}$ operators 
which are, on the other hand, supposed to be responsible for the main part of the dynamic correlation and which are calculated from the usual projective CCSD amplitude equations
analogous to the single-reference CC method
\begin{subequations}
\begin{align}
  \label{eq:TCCSD_eq}
  \langle \Phi_i^a | H e^{T_\text{ext}} e^{T_\text{CAS}} | \Phi_0 \rangle_c &= 0                       &\{i,a\} \not\subset \text{CAS}, \\
  \langle \Phi_{ij}^{ab} | H e^{T_\text{ext}} e^{T_\text{CAS}} | \Phi_0 \rangle_c &= 0  &\{i,j,a,b\} \not\subset \text{CAS}.
\end{align}
\end{subequations}

The TCC approach has been successfully applied \cite{tcc-appl,cyclobut-tailored-2011} and generally performs well,
although a large active space and CASSCF orbitals might be required for good accuracy \cite{kinoshita_2005}.
TCC also features the desirable property of being rigorously size-extensive \cite{kinoshita_2005}.

In order to circumvent the prohibitive scaling of the CASCI method, when large active spaces are used, we propose to use MPS wave functions generated by the DMRG method to acquire active space amplitudes, that correctly describe the non-dynamic correlation in the subsequent CCSD calculations.

The DMRG method
\cite{schollwock_2005} is a variational procedure which optimizes the wave function in the form of MPS \cite{schollwock_2011}.
It is a non-linear wave function ansatz made from the product of variational objects (matrices) corresponding to each site of a one-dimensional lattice which in QC represents a chain of molecular orbitals.
Therefore MPS 
refers to the wave function ansatz, whereas DMRG to the efficient self-consistent optimization algorithm which provides it. 
In QC version of DMRG (QC-DMRG) \cite{Legeza-2008,marti_2010,chan_review,wouters_review,legeza_review,yanai_review}, correlations between individual molecular orbitals are taken into account by means of an iterative procedure that variationally minimizes the energy of the electronic Hamiltonian.
The method eventually converges to the FCI solution in a given orbital space, i.e. to CASCI.

The practical version of DMRG is the two-site algorithm, which, in contrast to the one-site approach,
is less prone to get stuck in local minimum \cite{schollwock_2005}. It provides the wave function in the two-site MPS form \cite{schollwock_2011}

\begin{eqnarray}
  \label{eq:MPS_2site}
  | \Psi_\text{MPS} \rangle = \sum_{\{\alpha\}} \mathbf{A}^{\alpha_1} \mathbf{A}^{\alpha_2} \cdots \mathbf{W}^{\alpha_i \alpha_{i+1}} \cdots \mathbf{A}^{\alpha_n}| \alpha_1 \alpha_2 \cdots \alpha_n \rangle, \nonumber \\
\end{eqnarray}

\noindent
where $\alpha_i \in \{ | 0 \rangle, | \downarrow \rangle, | \uparrow \rangle, | \downarrow \uparrow \rangle \}$ and 
for a given pair of adjacent indices $[i, (i+1)]$, $\mathbf{W}$ is a four index tensor, which corresponds to the eigenfunction of the electronic Hamiltonian expanded in the tensor product space of four
tensor spaces defined on an ordered orbital chain, so called \textit{left block} ($M_l$ dimensional tensor space)  , \textit{left site} (four dimensional tensor space of $i^{\text{th}}$ orbital), \textit{right site} (four dimensional tensor space of $(i+1)^{\text{th}}$ orbital), and \textit{right block} ($M_r$ dimensional tensor space). 
The MPS matrices $\mathbf{A}$ are obtained by successive application of the singular value decomposition (SVD) with truncation on $\mathbf{W}$'s and iterative optimization by 
going through the ordered orbital chain from \textit{left} to \textit{right} and then sweeping back and forth.

The maximum dimension of MPS matrices which is required for a given accuracy, so called bond dimension
$[M_{\text{max}} = \text{max}(M_l, M_r)]$,
can be regarded as a function of the level of entanglement in the studied system \cite{legeza_2003b}. 
Among others, $M_{\text{max}}$ strongly depends on the order of orbitals along the one-dimensional chain \cite{legeza_2003a, moritz_2005} as well as their type \cite{fertitta_2014, krumnow_2015, amaya_2015}.

The two crucial correlation measures, which play an important role in tuning the performance of DMRG (e.g. employed in orbital ordering optimization), are single-orbital entanglement entropy ($s_i$) and mutual information ($I_{ij}$) \cite{legeza_2003b,legeza_2006,rissler_2006,barcza_2011}. $s_i$ quantifies the importance of orbital $i$ in the wave function expansion and can be computed as
$-\text{Tr} \rho_i \text{ln}  \rho_i$,
where $\rho_i$ represents the reduced density matrix of orbital $i$ \cite{legeza_2003b,Boguslawski-2012b,boguslawski_2013,barcza_2015}. Similarly, when substituting a single orbital by a pair of orbitals ($i$, $j$), the two-orbital entanglement entropy, $s_{ij}$, can be obtained. The mutual information then reads
$I_{ij} = s_{ij} - s_{i} - s_{j}$
and it describes how orbitals $i$ and $j$ are correlated with each other as they are embedded in the whole system \cite{rissler_2006,barcza_2011}.

When employing the two-site MPS wave function (Eq. \ref{eq:MPS_2site}) for the purposes of the TCCSD method,
the CI expansion coefficients $c_i^a$ and $c_{ij}^{ab}$ for $a,b,i,j \in \text{CAS}$ can be efficiently calculated by contractions of MPS matrices \cite{moritz_2007, boguslawski_2011}. 
We would like to note that using the two-site DMRG approach in practice means using the wave-function calculated at different sites and it can only be employed together with the dynamical block state selection (DBSS) procedure \cite{legeza_2003a} assuring the same accuracy along the sweep. Alternatively, one can use the one-site approach in the last sweep \cite{Zgid-2008b}.

Regarding the computational scaling of the DMRG-TCCSD method, it is indeed an interplay of contributions from both parent methods. The formal computational scaling of DMRG is $\mathcal{O}(M^3n^3) + \mathcal{O}(M^2n^4)$ \cite{white_1999}, where $M$ denotes the bond dimension and $n$ the number of orbitals in the DMRG space
\footnote{The term $\mathcal{O}(M^3n^3)$ is usually the dominant one.}. 
The scaling of CCSD \cite{gauss-encyclop} is $\mathcal{O}(N_{\text{occ}}^2N_{\text{virt}}^4) \approx \mathcal{O}(N^6)$, where to distinguish between DMRG and CC orbital spaces, $N$ refers to the size of the full (CC) orbital space and $n<N$. 
Which contribution is in practice the rate limiting step depends on  
the size of the DMRG active space, the underlying entanglement, as well as the size of the system. 
Computation of the CI expansion coefficients itself is negligible compared to the cost of DMRG 
\footnote{It might seem at the first sight that it will scale as $\mathcal{O}(M^2n^5)$ [$\mathcal{O}(n^4)$ CI coefficients each requiring $\mathcal{O}(M^2 n)$ operations when contracting the MPS matrices].
However, when the formation of two-index intermediates and
construction of different terms during the whole sweep is used in the
same way as for the construction of DMRG two-body reduced density matrices \cite{Zgid-2008b},
the scaling can be reduced to $\mathcal{O}(M^2n^4)$.}.

Looking upon at Eq.~(\ref{eq:TCCSD}) and taking into account that the action of $\big(\exp(T^{(1)}_{\text{CAS}})+ \exp(T^{(2)}_{\text{CAS}})\big)$ (with the exact amplitudes) on the reference function $|\Phi_0\rangle$ \textit{approximates} the MPS wave function ($\Psi_{\text{MPS}}^{\text{(CCSD)}}$) \footnote{Even though this approximation may not be perfect, the method performs very well, as shown on the examples, due to the fact that energy expressed in intermediate normalization requires only T1 and T2 amplitudes and the active space ones are very accurate, indeed.}, the method can be viewed as an approximate coupled cluster ansatz with the MPS reference function
\begin{equation}
	| \Psi_{\text{DMRG-TCCSD}} \rangle \approx e^{\big(T^{(1)}_\text{ext} + T^{(2)}_\text{ext}\big)} | \Psi_{\text{MPS}}^{\text{(CCSD)}} \rangle.
\end{equation}
However, it uses a single Slater determinant as a Fermi vacuum which
introduces a certain bias, which might deteriorate the performance of
the method in exactly degenerate situations. In such cases, the method
will break the spatial symmetry of the degenerate components. 
For the same reason, the TCCSD method despite being size-extensive does not fulfill the size-consistency exactly \footnote{The underlying DMRG is size-consistent, provided that the active space has been properly chosen to include orbitals to which bonds dissociate in the fragments (analogous situation to CASSCF.}, however, with growing size of the active space the error will decrease and in the limit of including all
orbitals, the error must vanish, since TCCSD then becomes identical to FCI.
We have performed a test of size-consistency of the method by comparison of the energy of N$_2$ dimer separated by 100 a.u. with respect to double of the N atom energy. While standard (closed shell) CCSD fails to converge at all for the separated dimer,
TCCSD(6,6) has an error of 20.7 kcal/mol, which monotonically decreases
with the size of the active space, yielding 20.0 kcal/mol, 14.7 kcal/mol
and 13.4 kcal/mol for the spaces (10,14), (10,16), and (10,18), respectively. 
The advantage of employing DMRG together with TCC is that it enables using of large active spaces (up to 40 orbitlas in generic cases), decreasing 
significantly any such errors.            
Even if not suppressed completely, we still believe that the TCCSD method can be very useful in computational scenarios where size-consistency is not a critical issue, as demonstrated on our numerical examples.

In what follows, we denote the DMRG-TCCSD method by the abbreviation TCCSD($e$,$o$), where the numbers inside the brackets specify the DMRG active space, namely $e$ refers to the number of electrons and $o$ to the number of orbitals.

The chromium dimer (Cr$_2$) has been known for a long time as a particularly challenging small system in quantum chemistry. 
In order to adequately describe its intricate dissociation curve, the used method has to provide the best possible treatment of both non-dynamic and dynamic correlation. 
Over the decades, the problem has been tackled by many groups \cite{angeli_2006,goodgame_1985,andersson_1994,bauschlicher_1994,roos_1995,stoll_1996,dachsel_1999,roos_2003,celani_2004,zgid_2009,muller_2009,amaya_2015,kurashige_2011,reiher_cr2}.
Our aim was not to calculate the whole dissociation curve, 
but rather test the DMRG-TCCSD method on a single-point energy calculation for which the large-scale DMRG extrapolated energy has recently been published \cite{amaya_2015}. These results are considered as a FCI benchmark. 

Following \cite{amaya_2015}, we performed a single-point calculation, with the chromium atoms being placed 1.5 \AA{} apart. 
According to the single-orbital entanglement entropy profiles, 
we have chosen three active spaces: CAS(12,12) for $s_i > 0.2$; CAS(12,19) for $s_i > 0.05$ and CAS(12,21) for $s_i$ just under the 0.05 (after these two orbitals, a 
drop in $s_i$ values was observed).
The first CAS includes all the valence orbitals (4$s$ and 3$d$), the other spaces are augmented by double-shell orbitals. In particular, CAS(12,19) adds two 5$s$ and five 4$d$ orbitals and CAS(12,21) adds another two 4$d$ orbitals.

\begin{figure}[!ht]
  \centering
  \includegraphics[width=\linewidth]{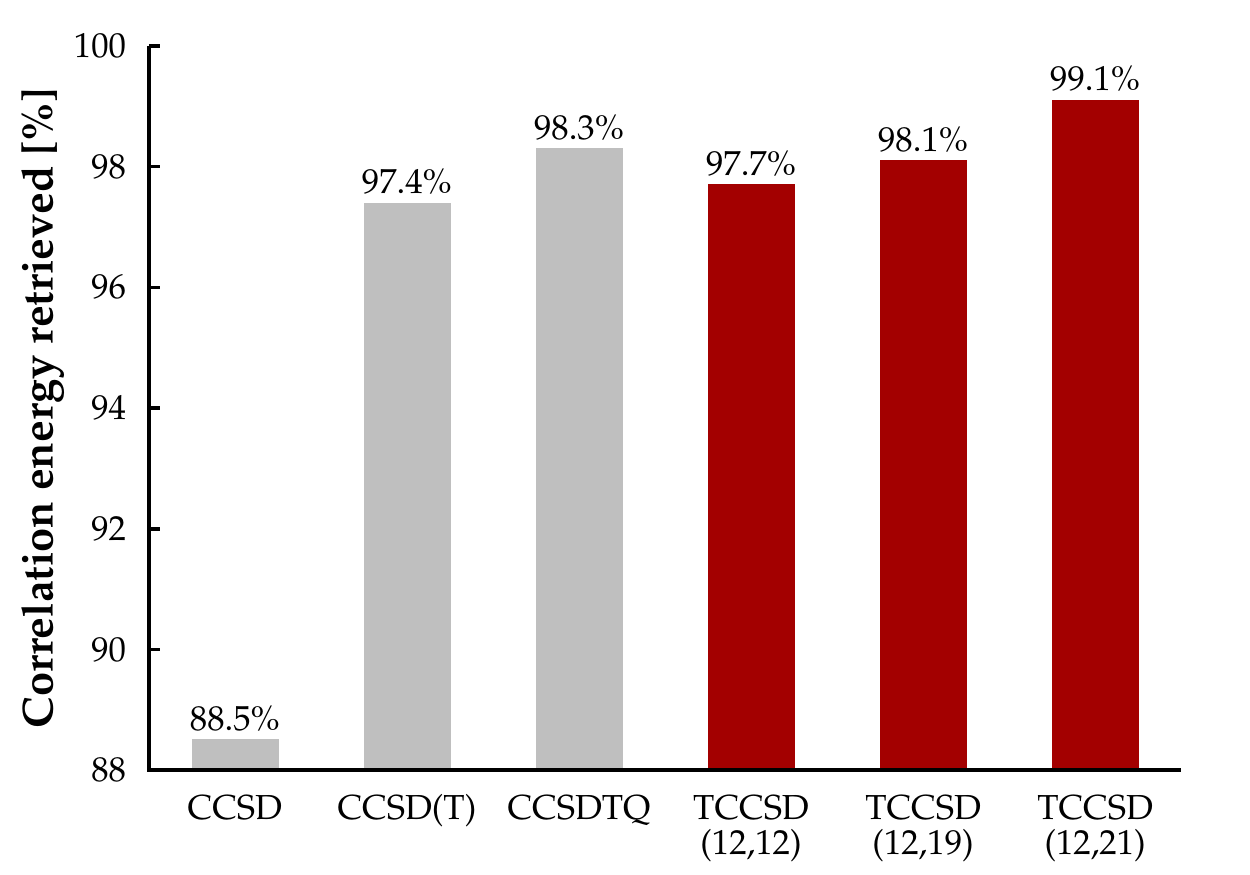}
  \caption{The amount of correlation energy retrieved (with respect to the extrapolated DMRG energies \cite{amaya_2015}) by TCCSD for the Cr$_2$ molecule ($r=1.5$ \AA) with SV basis. CCSD, CCSD(T) and CCSDTQ \cite{amaya_2015} energies are shown for comparison. \label{fig:cr2_bar}}
\end{figure}

The resulting DMRG, TCCSD and for comparison also CCSD, CCSD(T), and CCSDTQ \cite{amaya_2015} energies are shown in Table \ref{tab:cr2_energies}. 
The amount of retrieved correlation energy (with respect to the extrapolated DMRG energies \cite{amaya_2015}) for CC and TCCSD methods is plotted in Figure \ref{fig:cr2_bar}. 
The calculations systematically ameliorate with augmenting active space and in case of the largest TCCSD(12,21) calculation, we were able to retrieve more than 99\% of the overall correlation energy. This is a significant improvement upon stand-alone CCSD or DMRG calculations and it even surpasses the considerably more demanding CCSDTQ method.

\begin{table}[!ht]
\centering
\caption{The TCCSD energies ($E+2086$ in a.u.) of the Cr$_2$ molecule ($r=1.5$ \AA) with SV basis for different active spaces, with their respective DMRG ($\chi = 10^{-5}$)\footnote{For the definition of $\chi$, see the Computational details.} energies. CCSD, CCSD(T), CCSDTQ energies are shown for comparison.}
\renewcommand{\arraystretch}{1.2}
\label{tab:cr2_energies}
\begin{tabular}{cc}
\hline \hline
\parbox[c]{0.23\textwidth}{Method}       & \parbox[c]{0.23\textwidth}{$E+2086$}    \\ \hline
DMRG(12,12)  & $-0.071746$ \\
TCCSD(12,12) & $-0.424826$ \\
DMRG(12,19)  & $-0.228125$ \\
TCCSD(12,19) & $-0.428037$ \\
DMRG(12,21)  & $-0.252552$ \\
TCCSD(12,21) & $-0.437171$ \\ \hline
CCSD         & $-0.344277$ \\
CCSD(T) \cite{amaya_2015}      & $-0.422229$ \\
CCSDTQ \cite{amaya_2015}      & $-0.430244$ \\ \hline
DMRG(48,42)\footnote{Extrapolated DMRG energies serving as a FCI benchmark.} \cite{amaya_2015}          & $-0.444784$ \\ \hline \hline
\end{tabular}
\end{table}

The next system we have chosen for tests of the DMRG-TCCSD method is the nitrogen molecule (N$_2$).
It is well known that a proper description of the triple bond breaking process in N$_2$ requires reliable multireference treatment. 
For example the single-reference CCSD method fails by predicting an unphysical hump on the potential energy surface (PES) of the $\text{X}^1 \Sigma ^+ _\text{g}$ electronic state for about twice the equilibrium distance. On the other hand, as has been shown by Kinoshita \textit{et al.} \cite{kinoshita_2005} 
already TCCSD(6,6) corrects this unphysical behavior [see Fig. 3 of Supporting Information (SI)].

In order to accurately calculate spectroscopic parameters like vibrational frequencies ($\omega_e$) or anharmonicities ($\omega_e x_e$), a 
high quality PES is required, which makes them good tests
for the DMRG-TCCSD method.

As in the previous example, the DMRG active space was selected according to the single-orbital entanglement entropy values. 
We have selected 19 orbitals which complied with $s_i > 0.02$.
The final DMRG(10,19) and TCCSD(10,19) results (vibrational frequencies, anharmonicities, and equilibrium bond lengths) together with the single-reference CCSD and TCCSD(6,6) results are shown in
Table \ref{tab:n2_spect_par}.

\begin{table}[!ht]
\centering
\caption{Spectroscopic parameters of the $\text{X}^1 \Sigma ^+ _\text{g}$ electronic state of N$_2$ calculated with the cc-pVTZ basis together with the experimental values taken from \cite{lofthus_1977}. Vibrational frequencies ($\omega_e$) and anharmonicities ($\omega_ex_e$) are shown in cm$^{-1}$, bond lengths in \AA. Absolute values of deviations from the experimental results are also displayed.}  

\renewcommand{\arraystretch}{1.2}
\label{tab:n2_spect_par}
\begin{tabular*}{\linewidth}{l @{\extracolsep{\fill}} ccccccc}
\hline \hline
             & $\omega_e$ & $|\Delta \omega_e|$ & $\omega_ex_e$ & $|\Delta \omega_ex_e|$ & $r_0$ & $|\Delta r_0|$ \\
\hline
CCSD         & 2423.3   & 64.7  & 12.75    & 1.57    & 1.0967   & 0.0010  \\
TCCSD(6,6)   & 2376.3   & 17.7  & 13.57    & 0.75    & 1.1009   & 0.0032  \\
DMRG(10,19)  & 2298.8   & 59.8  & 13.72    & 0.60    & 1.1112   & 0.0135  \\
TCCSD(10,19) & 2347.3   & 11.3  & 13.91    & 0.41    & 1.1036   & 0.0059  \\ \hline
Experiment & 2358.57  &  & 14.324     &   & 1.09768     \\ \hline \hline       
\normalsize
  \end{tabular*}
\end{table}

As can be seen, the TCCSD(10,19) method gives the best agreement with the experimental vibrational frequencies and anharmonicities, improving the CCSD and DMRG(10,19) vibrational frequencies by more than 53 cm$^{-1}$ and 48 cm$^{-1}$ respectively and anharmonicities by more than 1.1 cm$^{-1}$ and 0.2 cm$^{-1}$ respectively. It also gives the vibrational frequencies and anharmonicities superior to TCCSD(6,6), by more than 6 cm$^{-1}$ in case of the vibrational frequency and 0.3 cm$^{-1}$ for the anharmonicity. 
Only the TCCSD(10,19) equilibrium bond length is slightly worse than the CCSD value, which is however justifiable, as the CCSD methods works well around the energy minimum where the wave function exhibits a single-reference nature. 
Nevertheless, the error of 0.006 \AA~ for the TCCSD(10,19) equilibrium bond length represents a fairly good accuracy.
The TCCSD(6,6) dissociation energy ($D_e$) computed as difference of the N$_2$ energy at the optimum geometry and the double of N atom energy equals 213.7 kcal/mol and lies 11 kcal/mol under the experimental value ($D_e^{\text{exp}} = 225$ kcal/mol \cite{lofthus_1977}). It improves the CASSCF(6,6) dissociation energy by 10 kcal/mol and the CCSD dissociation energy by 6.5 kcal/mol ($D_e^{\text{CASSCF(6,6)}} = 203.8$ kcal/mol, $D_e^{\text{CCSD}} = 207.2$ kcal/mol).

The last system which we have computed is oxo-Mn(Salen). It catalyzes the enantioselective epoxidation of unfunctional olefins \cite{jacobsen_1990, katsuki_1990} and it has been studied extensively with different multireference methods  \cite{ivanic_2004, sears_2006, ma_2011}, most recently also with the DMRG methodology \cite{wouters_2014, amaya_2015, stein_2016}. 
Despite huge efforts, the energetic ordering of the lowest singlet and triplet states is still not clear and proper answer requires studies of the effect of dynamic correlation.
The ordering of the lowest spin states is an important issue indeed, since different reaction paths have been suggested depending on the spin state \cite{abashkin_2001}.
To the best of our knowledge, we report the first ``post-DMRG'' computations of this system. 

\begin{figure*}[!ht]
    \centering
    \begin{subfigure}[b]{0.5\textwidth}
        \centering
        \includegraphics[width=\linewidth]{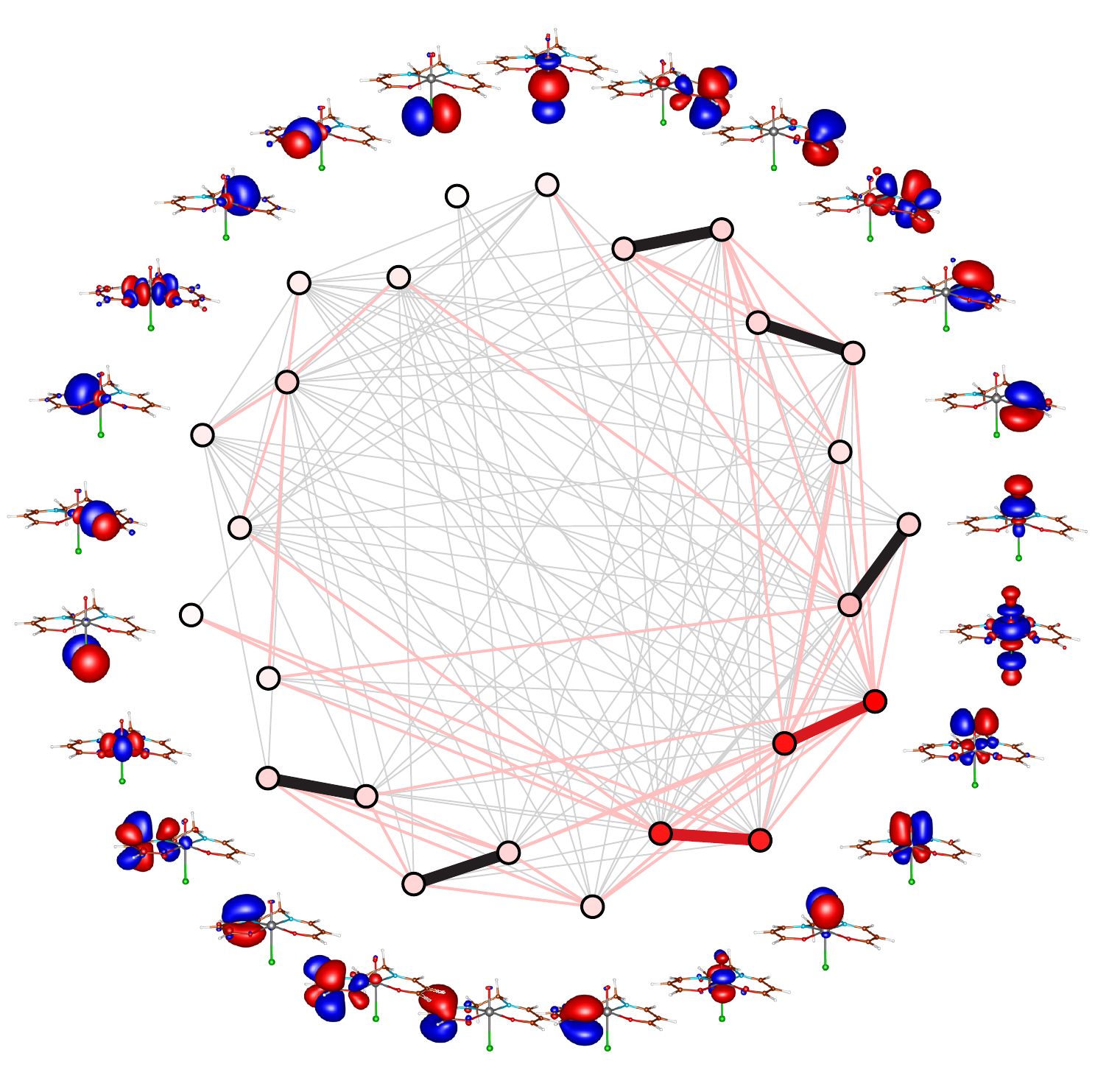}
        \caption{$^1A$ state}
    \end{subfigure}%
    ~ 
    \begin{subfigure}[b]{0.5\textwidth}
        \centering
        \includegraphics[width=\linewidth]{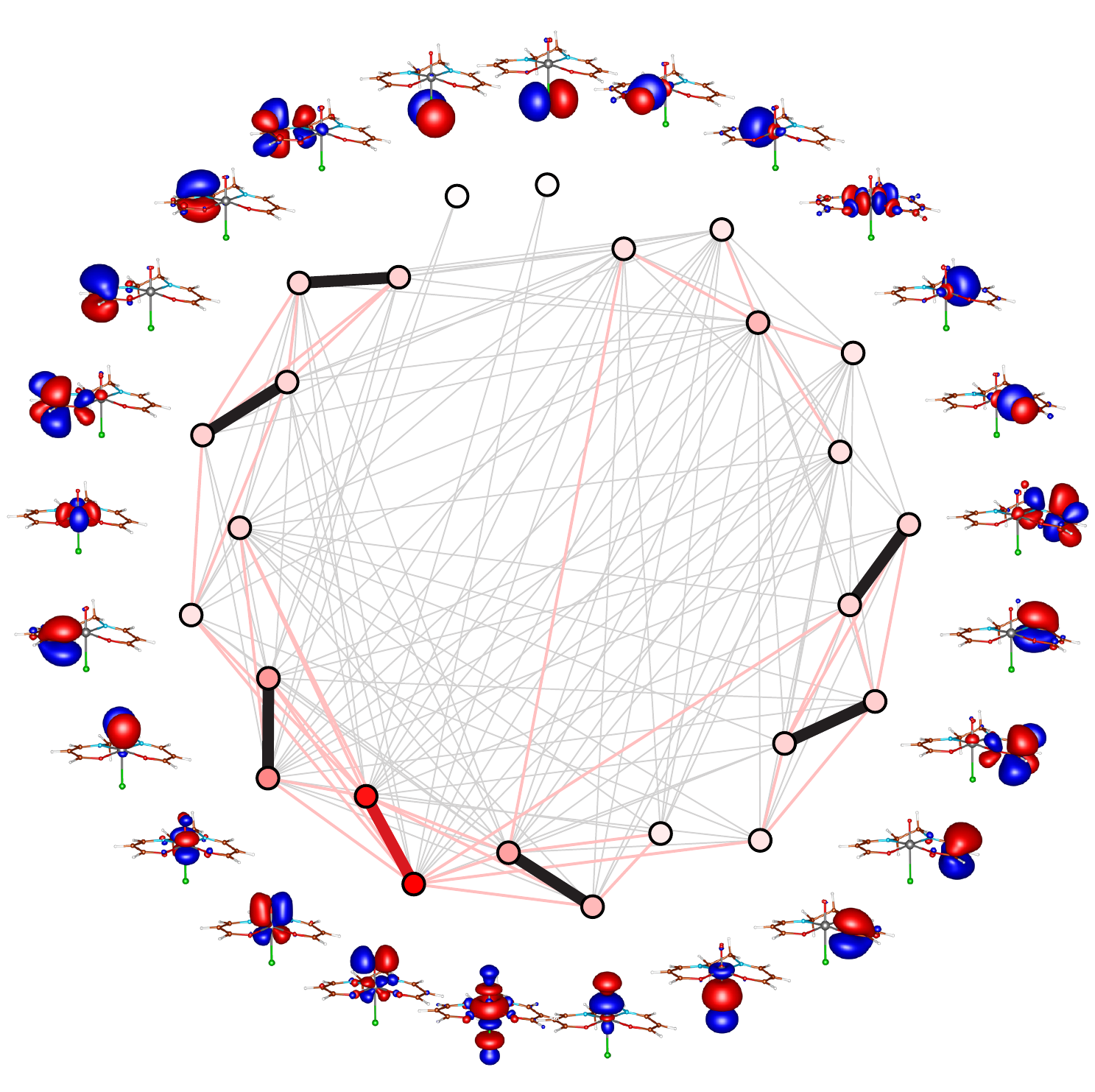}
        \caption{$^3A$ state}
    \end{subfigure}
    \caption{The CAS split-localized orbitals and their mutual information ($M=512$) for $^1A$ and $^3A$ states of oxo-Mn(Salen) with 6-31G$^*$ basis. The mutual information is color-coded: the thick red lines correspond to the strongest correlations (order of magnitude $1$), followed by black ($10^{-1}$), pink ($10^{-2}$) and grey ($10^{-3}$). One-site entropy values are represented by color gradient of the respective dot: red being the largest value and white being zero.}
  \label{oxo_salen_mutinfo}
\end{figure*}

In case of oxo-Mn(Salen), we followed the work of Olivares-Amaya et al.\cite{amaya_2015} in selection of the active space. The active space contained: 5 Mn 3$d$ orbitals, 10 $\pi$ orbitals of the equatorial conjugated rings (C, N, O atoms), 4 equatorial 2$p$ orbitals forming Mn-N and Mn-O $\sigma$ bonds, 3 2$p$ orbitals for axial O as well as Cl atoms, which resulted in CAS(34,25).
The split-localized molecular orbitals forming the DMRG active space with their respective mutual information are presented in Figure \ref{oxo_salen_mutinfo}. Our TCCSD and DMRG $^1A$ and $^3A$ energies together with previous DMRG and DMRG-SCF results are listed in Table \ref{tab:oxo_salen}. As can be seen, our DMRG(34,25) results agree with the DMRG-SCF results of Wouters et al. \cite{wouters_2014} in predicting the $^3A$ state to be the ground state\footnote{At our first attempt, we perfectly reproduced the DMRG results of Olivares-Amaya et al.\cite{amaya_2015} ($\Delta E$ differed by less than 0.1 kcal/mol, which is because our CAS was bigger by one occupied orbital), however we found out that we did not converge to the global $^3A$ state ROHF minimum, resulting in a wrong characteristic excitation of the $^3A$ state (it is unfortunately not mentioned in Ref. \cite{amaya_2015}). 
The correct character of the $^3A$ state which corresponds to the global ROHF minimum is shown in Fig. 7 of SI. 
The TCCSD method, nevertheless, predicted the $^3A$ state to be the ground state even in this case with $\Delta E = -0.4$ kcal/mol.}.
In our case, the singlet-triplet gap is higher in absolute value, which can be assigned to the fact that we did not optimize the orbitals. However, inclusion of the dynamic correlation through the TCCSD approach decreases the gap, 
suggesting that the $^3A$ state is lower in energy than the $^1A$ state by $3.6$ kcal/mol.

\begin{table}[!ht]
\centering
\caption{The TCCSD $^1A$ and $^3A$ energies ($E+2251$ in a.u.) and energy differences [$\Delta E = E(^3A) - E(^1A) $ in kcal/mol] of oxo-Mn(Salen) with 6-31G$^*$ basis. DMRG and previous DMRG and DMRG-SCF results are shown for comparison.}
\renewcommand{\arraystretch}{1.2}
\label{tab:oxo_salen}
\begin{tabular}{lrrr}
\hline \hline            
\parbox[c]{0.2\textwidth}{Method}             & \parbox[c]{0.1\textwidth}{$E(^1A)$}         & \parbox[c]{0.1\textwidth}{$E(^3A)$}        & \parbox[c]{0.05\textwidth}{$\Delta E$} \\ 
\hline
DMRG-SCF(28,22) \cite{wouters_2014}    & $-0.5498$\phantom{00}       & $-0.5578$\phantom{00}      & $-5.0$                             \\ 
DMRG(32,24) \cite{amaya_2015}        & $-0.304712$     & $-0.304128$    & ~$0.4$                              \\ 
DMRG-SCF(26,21)\footnote{The cc-pVDZ basis set was employed.} \cite{stein_2016}        & $-0.796326$     & $-0.795396$    & $~0.6$   \\ 
DMRG(34,25)        & $-0.410926$     & $-0.431039$    & $-12.6$                            \\
TCCSD(34,25)       & $-2.727314$     & $-2.733037$    & $-3.6$                             \\
\hline
\hline
\end{tabular}
\end{table}

In this letter, we have presented a novel method for accurate treatment of strongly correlated molecules which, in the spirit of TCC \cite{kinoshita_2005,bartlett_mrcc_2012,tcc-appl,cyclobut-tailored-2011}, combines the CC theory, in particular CCSD, with the DMRG method. In this approach, DMRG is responsible for a proper description of the non-dynamic electron correlation and CCSD is supposed to deal with the major part of the remaining dynamic correlation. 

In spite of being conceptually simple, the first results of the benchmark calculations on the Cr$_2$, N$_2$, and oxo-Mn(Salen) molecules are indeed very encouraging. In case of Cr$_2$ ($r=1.5$ \AA), for which the extrapolated DMRG energy is available,
we were able to recover more than $99~\%$ of the correlation energy with the TCCSD(12,21) method, compared to $88.5~\%$ of the standard CCSD method.
Regarding the N$_2$ example, with the TCCSD(10,19) method, we were able to obtain the vibrational frequency and the equilibrium bond length with errors about $0.5~\%$.
More importantly, this method provided the anharmonicity, which is more sensitive to the shape of the potential energy curve further from the energy minimum, with an error less than $3~\%$, compared to $11~\%$ error of the standard CCSD method. 
In case of oxo-Mn(Salen), we have presented the first ``post-DMRG'' calculations whose aim was to shed more light on the energy ordering of the lowest spin states. Our results are in agreement with the results of Wouters et al. \cite{wouters_2014}, predicting the triplet state to be 3.6 kcal/mol lower than the singlet one.

The DMRG-TCCSD method in fact represents the simplest version of DMRG-externally corrected CC approaches. Alternative (and potentionally even more accurate) method which deserves future investigations is the reduced multireference CCSD method \cite{paldus-externalcorr,paldus-externalcorr2,paldus-externalcorr-new1,paldus-externalcorr-new2} employing the DMRG connected triples and quadruples in the active space.

\section*{Computational details \label{sec:computational}}

  We have extended the Budapest QC-DMRG code \cite{budapest_qcdmrg} for computations of active space coupled cluster amplitudes and interfaced it with the Orca program system \cite{orca}, in which we have implemented the TCCSD method.

  In all the production DMRG calculations, we have employed the dynamical block state selection (DBSS) procedure \cite{legeza_2003a,legeza_2004} with the truncation criterion set on entropy $\chi = S_{\text{max}} - S$, where $S_{\text{max}}$ denotes the entanglement entropy of the augmented block before the truncation and $S$ the truncated one.
  We have tested the effect of the truncation error $\chi$ on the final TCCSD  N$_2$ spectroscopic parameters and found that $\chi = 10^{-5}$ is sufficient for the presented accuracy.
This truncation criterion was used throughout the work 
and it
resulted in bond dimensions varying in the range of 1000-6000.
Note that $\chi$ is a tighter criterion than the more common $\delta \epsilon_{\text{TR}} = 1 - \sum{\lambda_i^2}$ with $\lambda_i$ being the Schmidt values, in our case by almost two orders of magnitude.
  The orbitals for the DMRG active spaces were chosen according to their single-orbital entanglement entropies \cite{legeza_2003b, Boguslawski-2012b, stein_2016} calculated with fixed bond dimensions $M=512$ (see Figs. 2 and 5 of SI).
In case of N$_2$, the single-orbital entropies were averaged over 10 points, from which the spectroscopic parameters were computed.
  The Fiedler method \cite{barcza_2011,fertitta_2014} was used for optimization of the orbital ordering. 
  The DMRG runs were initialized using the CI-DEAS procedure \cite{legeza_2003b,legeza_review} and
  the energy convergence threshold measured between the two subsequent sweeps was set to $10^{-6}$ a.u.

  In case of Cr$_2$, the RHF orbitals computed in Ahlrichs' SV basis were used for the subsequent DMRG and TCCSD calculations (see Fig. 1 of SI).
  For N$_2$ 
we have used the CASSCF(6,6)/cc-pVTZ orbitals with the active space consisting of six 2$p$ orbitals in all DMRG and TCCSD calculations (see Fig. 4 of SI).
Likewise, we have excluded two 1$s$ orbitals from the TCCSD correlation treatment.
The N$_2$ spectroscopic parameters were calculated employing the
Dunham analysis \cite{dunham_1932}.
In case of oxo-Mn(Salen) we have used the singlet CASSCF(10,10)/6-31G$^*$ geometry of Ivanic et al. \cite{ivanic_2004} (see Fig. 6 of SI). As in Ref. \cite{amaya_2015}, we have employed the triplet 6-31G$^*$ ROHF orbitals which were for easier selection of the DMRG active space split-localized: for $^1A$ state all the valence and 32 virtual orbitals; for $^3A$ all the valence, two singly occupied and 31 virtual orbitals. Again, the core orbitals were excluded from correlation treatment.

\section*{Acknowledgment}

We would like to thank Ch. Krumnow, R. Schneider, and Sz. Szalay for helpful discussions and F. Wennmohs for the Orca technical assistance.
This work has been supported by the Czech Science Foundation (grant no. 16-12052S),
by the Czech Ministry of Education, Youth and Sports from the Large Infrastructures for Research, Experimental Development and Innovations project ``IT4Innovations National Supercomputing Center - LM2015070",
DAAD/16/07 project,
Hungarian-Czech Joint Research Project MTA/16/05, and
the Hungarian Research Fund (OTKA) (grants no. NN110360 and no. K120569). \\

L.V. and A.A. contributed equally to this work. \\

\textbf{Supporting Information Available:} 
DMRG active space orbitals for N$_2$ and Cr$_2$ examples with mutual information and single-orbital entanglement entropy plots, plot of N$_2$ PES, employed oxo-Mn(Salen) geometry, and oxo-Mn(Salen) orbitals involved in the characteristic singlet to triplet excitation.

\bibliography{dmrg_tcc,ors,dmrg,cc,references}
\bibliographystyle{achemso}

\end{document}